\documentclass[12pt]{article}
\usepackage{latexsym}
\usepackage{amsmath}
\usepackage{amssymb}
\usepackage{amsfonts}
\usepackage{comment}
\usepackage{graphicx}
\usepackage{hyperref}
\usepackage{verbatim}
\usepackage{enumerate,bbold}
\newcommand{\be}{\begin{equation}} \newcommand{\ee}{\end{equation}}
\newcommand{\bea}{\begin{eqnarray}} \newcommand{\eea}{\end{eqnarray}}

\newcommand{\calh}{{\mathcal H}}

\newcommand{\ket}{\rangle}

\begin{document}
\baselineskip=20pt

\begin{flushright} CALT-68-2929  \end{flushright}

\begin{center}
{\Large{\bf Many Worlds, the Born Rule, \\ and Self-Locating Uncertainty}}%
\footnote{A version of this paper appears as a chapter in {\it Quantum Theory: A Two-Time Success Story, Yakir Aharonov Festschrift} (2013), D.C. Struppa, J.M. Tollaksen, eds. (Springer-Verlag), p. 157. This work is a summary of a more comprehensive paper \cite{sc2013}. This is an updated version of the original published chapter.  Section \ref{mixed} is new.  The derivations in section \ref{sec5} have been revised (thanks to Jerry Finkelstein for alerting us to problems with our original treatment of these cases).}

\vspace*{0.1in}
Sean M.\ Carroll$^*$ and Charles T.\ Sebens$^\dagger$

\it $^*$Walter Burke Institute for Theoretical Physics, California Institute of Technology

\it $^\dagger$Philosophy Department, University of Michigan

 {\tt seancarroll@gmail.com}, {\tt csebens@gmail.com}
\vspace*{0.1in}
\end{center}

\begin{abstract}
We provide a derivation of the Born Rule in the context of the Everett (Many-Worlds) approach to quantum mechanics. Our argument is based on the idea of self-locating uncertainty: in the period between the wave function branching via decoherence and an observer registering the outcome of the measurement, that observer can know the state of the universe precisely without knowing which branch they are on. We show that there is a uniquely rational way to apportion credence in such cases, which leads directly to the Born Rule. Our analysis generalizes straightforwardly to cases of combined classical and quantum self-locating uncertainty, as in the cosmological multiverse.
\end{abstract}
\baselineskip=14pt

\section{Introduction}
\label{sec:1}

A longstanding puzzle in the Many-Worlds or Everett approach to quantum mechanics (EQM) \cite{everett,wallace2012} is the origin of the Born Rule:  the probability of finding a post-measurement system in an eigenstate $|a\rangle$ of an observable $A$, given that the system is prepared in state $|\psi\rangle$, is given by $|\langle a|\psi\rangle|^2$. Here we summarize and discuss the resolution of this problem that we recently developed \cite{sc2013}, in which the Born Rule is argued to be the uniquely rational way of dealing with the self-locating uncertainty that inevitably accompanies branching of the wave function.  A similar approach has been advocated by Vaidman \cite{vaidman2011}; our formal manipulations closely parallel those of Zurek \cite{zurek2005}.

Ours is certainly not the first attempt to derive the Born Rule within EQM. One approach is to show that, in the limit of many observations, branches that do not obey the Born Rule have vanishing measure \cite{hartle1968,Farhi:1989pm,Aguirre:2010rw}. A more recent twist is to use decision theory to argue that a rational agent should act as if the Born Rule is true \cite{deutsch1999, wallace2002,greaves2004, wallace2010b}. Another approach is to argue that the Born Rule is the only well-defined probability measure consistent with the symmetries of quantum mechanics \cite{gleason,zurek2005}. 

While all of these ideas have some degree of merit, they don't seem to have succeeded in convincing a majority of experts in the field. Our purpose here is not to criticize other approaches (there may be many valid ways to derive a correct answer), but to provide a simple and hopefully transparent alternative derivation that is physics-oriented while offering a clear answer to the question of how probabilities arise in EQM, a deterministic theory.

The main idea we use is that of \emph{self-locating uncertainty} \cite{lewis1979}: the condition of an observer who knows that the environment they experience occurs multiple times in the universe, but doesn't know which example they are actually experiencing. We argue that such a predicament inevitably occurs in EQM, during the ``post-measurement/pre-observation'' period between when the wave function branches due to decoherence (measurement) and when the observer registers the affect of the branching (observation). A naive analysis might indicate that, in such a situation, each branch should be given equal likelihood; here we demonstrate that a more careful treatment leads us inevitably to the Born Rule for probabilities.

\section{Everettian Quantum Mechanics}

In EQM, the quantum state is described by a vector $|\Psi\rangle$ in a Hilbert space $\calh$, evolving under the influence of a self-adjoint Hamiltonian $H$ according to Schr\"odinger's equation
\be
  H|\Psi\rangle = i\hbar\partial_t|\Psi\rangle.
\ee
This smooth unitary evolution is supposed to account for absolutely all the quantum dynamics; there is no separate rule governing ``wave function collapse.'' Rather, we model the observer as well as the system as part of the quantum state, and unitary evolution causes the state of the universe to split into multiple non-interacting branches, each associated with a possible measurement outcome.

Consider an example in which the system is a single qubit initially in a state $|\psi\rangle = (1/\sqrt{2})(|{\uparrow}\rangle + |{\downarrow}\rangle)$, where the arrow denotes the value of the spin along the $z$-axis. The observer is initially uncorrelated with the spin, in a ready state $|O_0\rangle$. The measurement process is described via the following form of unitary evolution:
\bea
  |\Psi\rangle &=& \frac{1}{\sqrt{2}} |O_0\rangle\left(|{\uparrow}\rangle + |{\downarrow}\rangle\right)\\
&\rightarrow&  \frac{1}{\sqrt{2}} \left(|O_\uparrow\rangle|{\uparrow}\rangle +  |O_\downarrow\rangle|{\downarrow}\rangle\right).\label{twopeople}
\eea
Here, $|O_\uparrow\rangle$ and $|O_\downarrow\rangle$ represent states in which the observer has measured spin-up and spin-down, respectively. The wave function has not collapsed, but the observer is now described by a superposition of different measurement outcomes. 

The first challenge for such an approach is obvious: in the real world, it never \emph{feels} like we are in a superposition of measurement outcomes. We see the spin up or down, Schr\"odinger's Cat alive or dead -- never a superposition of different possibilities. Everett's insight was that, if a measurement of a spin that was originally in either of the eigenstates $|{\uparrow}\rangle$ or $|{\downarrow}\rangle$ leaves the observer with the impression of a definite measurement outcome, then the linearity of quantum mechanics implies that a superposition of such states should lead to two definite experiences.  The wave function in equation (\ref{twopeople}) represents two agents seeing two different outcomes, not one agent somehow experiencing an indeterminate outcome.

This story becomes more plausible once decoherence is understood as a crucial part of quantum mechanics. In a realistic situation, the observer and system do not constitute the entire universe; there is also an environment, generally with many more degrees of freedom. Initially the environment, like the observer, is in a state $|\omega_0\rangle$
that is unentangled with the system under consideration. But if the system and the environment are allowed to interact -- as is practically inevitable if the system is a macroscopic object like Schr\"odinger's Cat, constantly radiating and breathing (or failing to) and so forth -- then entanglement with the environment quickly ensues (typically before entanglement with the observer):
\bea
  |\Psi\rangle &=& \frac{1}{\sqrt{2}} |O_0\rangle\left(|{\uparrow}\rangle + |{\downarrow}\rangle\right)|\omega_0\rangle \nonumber\\
&\rightarrow&  \frac{1}{\sqrt{2}} |O_0\rangle\left(|{\uparrow}\rangle|\omega_\uparrow\rangle
 + |{\downarrow}\rangle|\omega_\downarrow\rangle\right).
 \label{decoherence}
\eea
In a generic situation, the entangled environment states will be nearly orthogonal: $\langle\omega_\uparrow| \omega_\downarrow\rangle \approx 0$. In that case, the component describing the up spin will no longer be able to interfere with the component describing the down spin. We say that decoherence has occurred, and the wave function
has branched. Decoherence helps explain how EQM is a theory of distinct causally well-isolated ``worlds.''\footnote{EQM is time-symmetric, but branching occurs toward the future, and not toward the past, because the low-entropy early universe was relatively free of entanglements between subsystems.}

A popular objection to EQM is that it is ontologically extravagant -- an incredible number of unobservable worlds are invoked to help explain observations within the single world to which we have access. This objection is misplaced. Any viable version of quantum mechanics involves a Hilbert space $\calh$ of very high dimensionality.  The holographic principle suggests that the dimensionality of the Hilbert space describing our observable universe is at least $\exp(10^{120})$ \cite{Banks:2000fe}, and there is good reason to believe it is infinite \cite{Carroll:2008yd,Boddy:2014eba}. In EQM, the size of Hilbert space remains fixed, but the state vector describes an increasing number of distinct worlds as it evolves. The potential for describing many worlds was always there; the objection that there are too many universes is really an objection that Hilbert space is too big, which would apply equally well to any approach to quantum mechanics which includes a state vector. A proper measure of ontological extravagance relies on the number of types of fundamental entities proposed by the theory (like the wave function) and laws that govern them, not the number of large scale structures (like quantum worlds) which emerge from them. EQM, which requires only a vector in a Hilbert space and a single evolution law, is ontologically quite restrained.

A more pressing concern is that the formalism of EQM offers little guidance to the preferred basis problem -- why do we collapse onto certain states and not others? We do not address this question in this paper, but there has been significant progress in understanding the origin of ``pointer states'' which arise from decoherence and are robust under macroscopic perturbations \cite{zurek81,jooszeh85}. Philosophically, there has been progress made in understanding how the many worlds of quantum mechanics can be emergent, arising dynamically from unitary evolution and not requiring the addition of new laws to govern their creation \cite{wallace2010c}.

Our concern here is with the origin of the Born Rule. In an equation such as (\ref{decoherence}), it is unclear what role the coefficients multiplying each branch should play for an observer living within the wave function. We will argue that they play a crucial role in justifying a probability calculus that leads us to the Born Rule. Along the way, we will see how probability can arise in a deterministic theory as agents evolve from perfect knowledge to self-locating uncertainty.

\section{Self-Locating Uncertainty}

Modern theories of cosmology often invoke ``large universes'' -- ones in which any given
local situation (such as a particular observer, in a particular macroscopic quantum state, with particular data about their surroundings) is likely to occur multiple times \cite{hartle2007, page2007, srednicki2010}. The setting could be something as dramatic as an inflationary multiverse, or as relatively pedestrian as an homogeneous cosmology with sufficiently large spatial sections. If the likely number of such duplicate observers is infinite, we face the cosmological measure problem. Even if it is finite, however, any one such observer finds themselves in a situation of self-locating uncertainty. They can know everything there is to know about the state of the universe and an arbitrary amount about their local environment, but still not be able to determine which such instantiation of that data they are experiencing.

This situation has been extensively studied in the philosophical literature (see {\it e.g.} \cite{bostrom2002, meacham2008, manleyF}), often in the case of hypothetical exact duplications of existing persons rather than large universes. One intuitively obvious principle for assigning probabilities in the face of such uncertainty is ``indifference,'' roughly: if all you know is that you are one of $N$ occurrences of a particular set of observer data, you should assign equal credence (a.k.a. ``degree of belief'' or just ``probability'') $1/N$ to each possibility. Elga \cite{elga2004} has given convincing arguments in favor of indifference in the case of identical classical observers. Crucially, this result is not simply postulated as the simplest approach to the problem, but rather derived from seemingly innocuous principles of rational reasoning. 

In EQM, self-locating uncertainty is inevitable: not with respect to different  locations in space, but with respect to different branches of the wave function. Consider again the branching process described in equation (\ref{decoherence}), but now we include an explicit measuring apparatus $A$ (which might represent an electron microscope, a Geiger counter, or other piece of experimental equipment). Imagine that we preserve quantum coherence in the system long enough to perform a measurement with the apparatus, which then (as a macroscopic object) rapidly becomes entangled with the environment and causes decoherence. Only then does the observer record the outcome of the measurement (normalizations have been omitted for convenience):
\bea
  |\Psi\rangle &=& |O_0\rangle\left(|{\uparrow}\rangle + |{\downarrow}\rangle\right)|A_0\rangle|\omega_0\rangle \\
  \ &\rightarrow& |O_0\rangle\left(|{\uparrow}\rangle|A_\uparrow\rangle 
  + |{\downarrow}\rangle|A_\downarrow\rangle\right)|\omega_0\rangle \\
  &\rightarrow&  |O_0\rangle\left(|{\uparrow}\rangle|A_\uparrow\rangle|\omega_\uparrow\rangle
 + |{\downarrow}\rangle|A_\downarrow\rangle|\omega_\downarrow\rangle\right)\\  
  &=&  |O_0\rangle|{\uparrow}\rangle|A_\uparrow\rangle|\omega_\uparrow\rangle
 + |O_0\rangle|{\downarrow}\rangle|A_\downarrow\rangle|\omega_\downarrow\rangle \label{sluqm}\\
  &\rightarrow&  |O_\uparrow\rangle|{\uparrow}\rangle|A_\uparrow\rangle|\omega_\uparrow\rangle
 + |O_\downarrow\rangle|{\downarrow}\rangle|A_\downarrow\rangle|\omega_\downarrow\rangle . 
\eea
Each line represents unitary time evolution except for (\ref{sluqm}), in which we have merely distributed the observer state for clarity. That is the moment we describe as post-measurement/ pre-observation. At that step, the wave function has branched -- decoherence has occurred, as indicated by the different environment states. The observer is still described by a unique state $|O_0\rangle$, but there are two copies, one in each branch. Such an observer (who by construction doesn't yet know the outcome of the measurement) is in a state of self-locating uncertainty.

A particularly clear example of such uncertainty would be a real-world version of the Schr\"odinger's Cat experiment. An actual cat interacts strongly with its environment and would not persist in a coherent superposition of alive and dead for very long; the wave function would branch long before a human experimenter opens the box. But in fact such uncertainty is generic. The timescale for decoherence for a macroscopic apparatus is extremely short, generally much less than $10^{-20}$\,sec. Even if we imagine an experimenter looking directly at a quantum system, the state of the experimenter's eyeballs would decohere that quickly. The timescale over which human perception occurs, however, is tens of milliseconds or longer. Even the most agile experimenter will experience \emph{some} period of self-locating uncertainty in which they don't know which of several branches they are on, even if it is too brief for them to notice.  Although the experimenter may not be quick-thinking enough to reason during this period, there are facts about what probabilities they ought to assign before they get the measurement data.

Naively, the combination of indifference over indistinguishable circumstances and self-locating uncertainty when wave functions branch is a disaster for EQM, rather than a way forward. Consider a case in which the amplitudes are unequal for two branches:
\be
  |\Psi\rangle = \sqrt{\frac{1}{3}}  |O_0\rangle|{\uparrow}\rangle|\omega_\uparrow\rangle
 + \sqrt{\frac{2}{3}}|O_0\rangle|{\downarrow}\rangle|\omega_\downarrow\rangle .\label{unequalamps}
\ee
The conditions of the two observers would seem to be indistinguishable from the inside; there is no way they can ``feel'' the influence of the amplitudes multiplying their branches of the wave function. Therefore, one might be tempted to conclude that Elga's principle of indifference implies that probabilities in EQM should be calculated
by \emph{branch-counting} rather than by the Born Rule -- every branch should be given equal weight, regardless of its amplitude.  In this case, equation (\ref{unequalamps}), that means assigning equal $50/50$ probability to up and down even though the branch weights are unequal. This would be empirically disastrous, as real quantum measurements don't work that way. We will now proceed to show why such reasoning is incorrect, and in fact a proper treatment of self-locating uncertainty leads directly to the empirically desirable conclusion. In Section~\ref{mixed} we generalize our result to cases where there is both classical and quantum self-locating uncertainty, as in the cosmological multiverse.

\section{The Epistemic Separability Principle}

We base our derivation of the Born Rule on what we call the \emph{Epistemic Separability Principle} (ESP), roughly: the probability assigned post-measurement/pre-observation to an outcome of an experiment performed on a specific system shouldn't depend on the physical state of other parts of the universe (for a more careful discussion see \cite{sc2013}). If I set out to measure the $z$-component of a spin in my laboratory, the probability of a particular outcome should be independent of the quantum state of some other spin in a laboratory on an alien planet around a distant star in the Andromeda galaxy. An essentially equivalent assumption is made by Elga in his discussion of classical self-locating uncertainty \cite{elga2004}.\footnote{The ESP is implicit in Elga's discussion of his TOSS \& DUPLICATION thought experiment, where he notes that the outcome of an additional coin toss should not affect the credence we assign to being either an original or a duplicated person with identical experiences.}  The ESP applies in both quantum and classical contexts.  In classical contexts, the ESP is compatible with Elga's indifference principle (see \cite{sc2013}).  In quantum contexts, it mandates the Born rule.  In EQM, the ESP amounts to the idea that the state of the environment shouldn't affect predictions that are purely about the observer/system Hilbert space.

Consider a Hilbert space that describes an observer, a system, and an environment:
 \be
   \calh = \calh_O \otimes \calh_S \otimes \calh_E .
 \ee
We consider general states of the universe, described by a state vector
\be
  |\Psi\ket = \sum_{a, i, \mu} \Psi_{a,i,\mu}|\psi_a\rangle |\phi_i\rangle |\omega_\mu\rangle,
\ee
where $\{\psi_a\}$, $\{\phi_i\}$, and $\{\omega_\mu\}$ are bases for the observer, system, and environment respectively, all of which are orthonormal: $\langle\psi_a|\psi_b\rangle = \delta_{ab}$ etc. Consider unitary transformations that act only on the environment, which we can write as
\be
  T = \mathbb{1}_{OS}\otimes U_E ,
\ee
where $U_E$ is a unitary matrix that acts on $\calh_E$. Then we can formulate the ESP in the context of EQM as the statement that probabilities in the observer/system subspace are unchanged by such transformations (here $s$ is a possible outcome of a measurement of the system $S$):
\be
  P(O {\rm \ measures\ } s|\Psi) = P(O {\rm \ measures\ } s|T[\Psi]). \label{PrincEpLoc}
\ee
In \cite{sc2013} we also offer a version of this principle using density matrices rather than directly in terms of transformations on states; roughly, the probability assigned to an outcome of an experiment depends only on the reduced density matrix of the observer/system subspace.\footnote{A possible worry is that by examining \emph{unitary} transformations on the environment we are secretly assuming some version of the Born Rule.  Unitary operators are distinguished by the fact that they preserve the inner product between any two states.  One might be concerned that this use of the inner product in characterizing unitary operators implicitly appeals to the Born rule.  It does not.  The inner product on Hilbert space is an essential part of the mathematical structure of quantum mechanics; what we are deriving is the connection between the inner product and probabilities.} The two formulations are equivalent if we are comparing states with identical Hilbert spaces for the environment.

ESP is an assumption about what factors are relevant to determining the probability of a particular experimental outcome in EQM post-measurement/pre-observation.   Transformations of the environment are considered irrelevant not because the system and observer are isolated from the environment -- the absence of interactions with the environment would make decoherence impossible -- but because a change to the environment that does not result in a change to the observer or to the system or to the relation of the observer to the system does not seem like it should matter for the project of locating oneself on a particular branch of the wave function (the project of assigning probabilities to possible measurement outcomes).

\section{Deriving the Born Rule}\label{sec5}

Consider a specific example where we have an observer, a spin with equal amplitudes
to be up or down, and an environment (again omitting the overall normalization):
\be
  |\Psi\rangle = |O\rangle |{\uparrow}\rangle |\omega_1\rangle +  |O\rangle |{\downarrow}\rangle |\omega_2\rangle.
  \label{psi}
\ee
Like equation (\ref{sluqm}), this is a state in which the observer has yet to observe the outcome and thus ought to be uncertain which branch they are on.  The environment states are assumed to be orthonormal (by decoherence), and without loss of generality we can take them to be the first two elements of an orthonormal basis $\{|\omega_\mu\rangle\}$.

We can write any environment unitary $U_E$ in the form
\be
  U_E = \sum_\mu |\tilde\omega_\mu\rangle\langle \omega_\mu|,
\ee
where the states $|\tilde \omega_\mu\rangle$ are another set of orthonormal vectors. We decompose the environment into a tensor product of two subsystems, one of which we will
label as a ``coin'' (although it could be of arbitrary dimension) and the other includes everything else:
\be
  \calh_E = \calh_C \otimes \calh_{\widehat E}.
\ee
Then we can construct an orthonormal basis $\{|\tilde\omega_\mu\rangle\}$ 
for the environment $\calh_E$ in which the first two basis vectors take the form
\bea
  |\tilde \omega_1\rangle &=& |{H}\rangle\otimes |\Omega_1\rangle,\\
  |\tilde \omega_2\rangle &=& |{T}\rangle\otimes |\Omega_2\rangle,  
\eea
where $|{H}\rangle$ (heads) and $|{T}\rangle$ (tails) are two orthonormal vectors in $\calh_C$, and
$|\Omega_1\rangle, |\Omega_2\rangle$ are orthonormal vectors in $\calh_{\widehat E}$. 
(Note that this is always possible, given our freedom to choose basis vectors in $\calh_E$, as long as that space is large enough and can be factorized as $\calh_C \otimes \calh_{\widehat E}$..)

Now we can construct two specific environment unitaries:
\bea
  U^{(1)}_E &=& \sum_{\mu} |\tilde\omega_\mu\rangle\langle\omega_\mu| ,\\
  U^{(2)}_E &=& |\tilde \omega_2\rangle\langle\omega_1| + |\tilde\omega_1\rangle\langle\omega_2|
  + \sum_{\mu > 2} |\tilde\omega_\mu\rangle\langle\omega_\mu| .
\eea
Acting on our state (\ref{psi}) we get
\bea
  |\Psi_1\rangle &\equiv&\left(\mathbb{1}_{OS}\otimes U_E^{(1)}\right)|\Psi\rangle \nonumber\\
  &=& |O\rangle |{\uparrow}\rangle |{H}\rangle|\Omega_1\rangle 
  +  |O\rangle |{\downarrow}\rangle |{T}\rangle|\Omega_2\rangle 
  \label{psi1} 
\eea
and
\bea
   |\Psi_2\rangle &\equiv&\left(\mathbb{1}_{OS}\otimes U_E^{(2)}\right)|\Psi\rangle\nonumber\\
  &=& |O\rangle |{\uparrow}\rangle |{T}\rangle|\Omega_2\rangle 
  +  |O\rangle |{\downarrow}\rangle |{H}\rangle|\Omega_1\rangle .
  \label{psi2}
\eea
In the $|\Psi_1\rangle$, the spin and the ``coin'' have become entangled so that the coin is heads if the particle was spin up, in $|\Psi_2\rangle$ the coin is heads if the particle was spin \emph{down}.

By the ESP, equation (\ref{PrincEpLoc}) -- the probability that the observer should assign to being on a branch where the measurement yielded spin up (or spin down) -- is equal in all of these states, since they are related by unitary transformations on the environment:
\bea
  P({\uparrow}|\Psi) = P({\uparrow}|\Psi_1) = P({\uparrow}|\Psi_2) ,\label{upprob}\\
  P({\downarrow}|\Psi) = P({\downarrow}|\Psi_1) = P({\downarrow}|\Psi_2) .\label{downprob}
\eea
However, we can also consider the coin to be ``the system," and the spin as part of the
environment. In that case, the two environments are related by a unitary transformation on the spin:
\be
  U_S = |{\uparrow}\rangle\langle{\downarrow}| + |{\downarrow}\rangle\langle{\uparrow}|.
\ee
Therefore, by analogous logic, the probability of heads or tails is equal in the two states $|\Psi_1\rangle$ and $|\Psi_2\rangle$:
\bea
  P({H}|\Psi_1) = P({H}|\Psi_2),\label{headprob}\\
  P({T}|\Psi_1) = P({T}|\Psi_2).\label{tailprob}
\eea

Looking at the specific states in (\ref{psi1}) and (\ref{psi2}), we notice that the branch of the wave function in which the coin is heads is the same as the one where the spin is up in $|\Psi_1\rangle$, but the one where the spin is down in $|\Psi_2\rangle$. So, in $|\Psi_1\rangle$ the particle is spin up (on the observer's branch) if and only if the coin is heads (on the observer's branch), and in $|\Psi_2\rangle$ the particle is spin down (on the observer's branch) if and only if the coin is heads (on the observer's branch). We therefore have
\bea
  P({\uparrow}|\Psi_1) = P({H}|\Psi_1),\label{oneprob}\\
  P({\downarrow}|\Psi_2) = P({H}|\Psi_2).\label{twoprob}
\eea
Comparing with (\ref{headprob}) we immediately get
\be
  P({\uparrow}|\Psi_1) = P({\downarrow}|\Psi_2),
\ee
and comparing that with (\ref{upprob}) and (\ref{downprob}) reveals
\be
    P({\uparrow}|\Psi) = P({\downarrow}|\Psi) = 1/2.
\ee
This is, of course, the result we expect from the Born Rule: when the components of the wave function have equal amplitudes, they get assigned equal probabilities. This shouldn't be surprising, as it is also what we would expect from naive branch-counting. However, notice that the equality of the amplitudes was crucially important, rather than merely incidental; had they not been equal, we would have been unable to fruitfully
compare results from different unitary transformations on the environment.

\begin{figure}[t]
\centerline{\includegraphics[width=10cm]{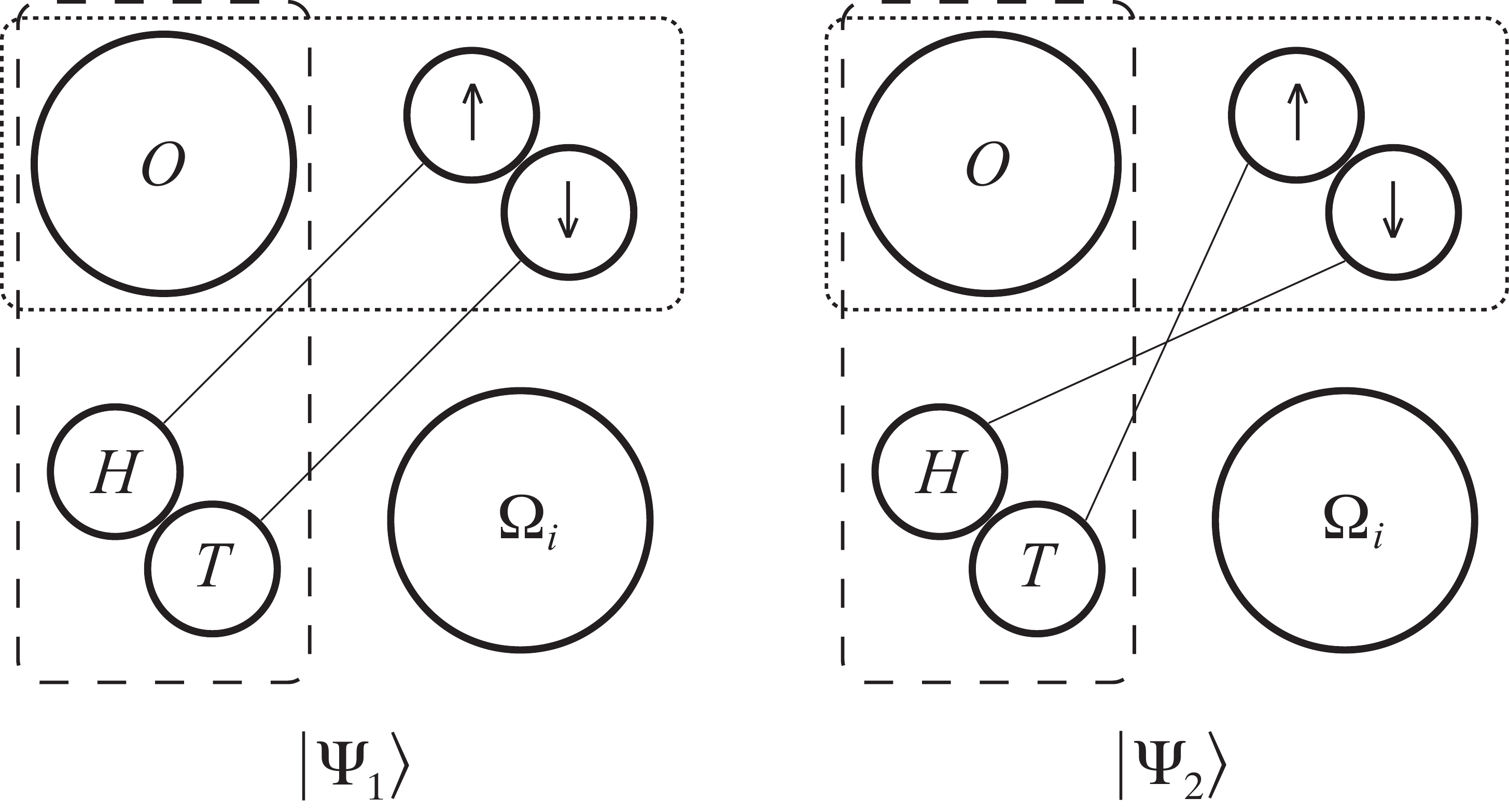}}
\caption{A schematic representation of the setup behind our derivation of the Born Rule. The states $|\Psi_1\rangle$ and $|\Psi_2\rangle$ are on the left and right, respectively. Factors denote the observer, the spin, the coin, and the rest of the environment. Thin diagonal lines connecting the spin and coin represent entanglement within different branches of the wave function. The horizontal/vertical boxes made from dotted/dashed lines show two different ways of carving out the ``Observer+System" subsystem from the ``Environment." The ESP implies that the probability of the system being in a particular state is independent of the state of the environment. Applying that rule to both the spin and coin systems implies the Born Rule as the uniquely rational way of assigning credences}
\label{pel1}
\end{figure}

It is therefore crucial to consider branches with unequal amplitudes. Here our logic follows that of Zurek \cite{zurek2005}. Start with a state where one branch has an amplitude greater than the other by a factor of $\sqrt{2}$:
\be
  |\Psi\rangle = |O\rangle |{\uparrow}\rangle |\omega_1\rangle +
  \sqrt{2}  |O\rangle |{\downarrow}\rangle |\omega_2\rangle.
  \label{psiu}
\ee
By ESP, the probabilities of spin up and spin down are unchanged if we transform the environment by
\begin{eqnarray}
U^{(0)}_E &=&|\widehat\omega_1\rangle\langle\omega_1| + \left(\frac{1}{\sqrt{2}}|\widehat\omega_2\rangle + \frac{1}{\sqrt{2}}|\widehat\omega_3\rangle\right)\langle\omega_2|
\nonumber\\
  &&+ \left(\frac{1}{\sqrt{2}}|\widehat\omega_2\rangle - \frac{1}{\sqrt{2}}|\widehat\omega_3\rangle\right)\langle\omega_3|+\sum_{\mu > 3} |\widehat\omega_\mu\rangle\langle\omega_\mu|,
\end{eqnarray}
where $\{|\widehat\omega_\mu\rangle\}$ is a new orthonormal basis for the environment. Then our state becomes
\be
  |\Psi_0\rangle = |O\rangle |{\uparrow}\rangle |\widehat\omega_1\rangle +
  |O\rangle |{\downarrow}\rangle |\widehat\omega_2\rangle+
  |O\rangle |{\downarrow}\rangle |\widehat\omega_3\rangle.
  \label{psiu2}
\ee
This reduces the problem of two branches with unequal amplitudes to that of three branches with equal amplitudes.

Following our previous logic, we construct a new orthonormal environment basis where the environment is decomposed into three factors: two systems whose states are unknown to the observer and a third system which includes everything else.  The first two factors each have three basis states (or any number of basis states, of which we will only care about three). Let the basis for the first factor be $\{|-\rangle, |0\rangle, |+\rangle\}$, and for the second factor we'll use card suits, $\{|\heartsuit\rangle, |\diamondsuit\rangle, |\spadesuit\rangle\}$. In terms of these we write another set of environment basis vectors $\{|\tilde \omega_\mu\rangle\}$ as:
\bea
  |\tilde \omega_1\rangle &=& |{+}\rangle\otimes |\heartsuit\rangle\otimes |\Omega_1\rangle,\\
  |\tilde \omega_2\rangle &=& |{-}\rangle\otimes |\diamondsuit\rangle\otimes |\Omega_2\rangle\\
  |\tilde \omega_3\rangle &=& |{0}\rangle\otimes |\spadesuit\rangle\otimes |\Omega_3\rangle,\\
  |\tilde \omega_4\rangle &=& |{-}\rangle\otimes |\spadesuit\rangle\otimes |\Omega_4\rangle\\
  |\tilde \omega_5\rangle &=& |{0}\rangle\otimes |\heartsuit\rangle\otimes |\Omega_5\rangle\\
  |\tilde \omega_6\rangle &=& |{+}\rangle\otimes |\diamondsuit\rangle\otimes |\Omega_6\rangle\\
  &\ldots&\nonumber
\eea
This list is not exhaustive, but that won't matter for our analysis.
Again we construct environment unitaries
\bea
  U^{(1)}_E &=& \sum_{\mu} |\tilde\omega_\mu\rangle\langle\widehat\omega_\mu| ,\\
  U^{(2)}_E &=& \sum_{\mu=1}^{3} |\tilde\omega_{\mu+3}\rangle\langle\widehat\omega_\mu|+\sum_{\mu=4}^{6} |\tilde\omega_{\mu-3}\rangle\langle\widehat\omega_\mu|+\sum_{\mu>6} |\tilde\omega_{\mu}\rangle\langle\widehat\omega_\mu| .\qquad
\eea
Acting on our state (\ref{psiu2}) we get
\be
  |\Psi_1\rangle 
  = |O\rangle |{\uparrow}\rangle |{+}\rangle|\heartsuit\rangle|\Omega_1\rangle 
  +  |O\rangle |{\downarrow}\rangle |{-}\rangle|\diamondsuit\rangle|\Omega_2\rangle 
  +  |O\rangle |{\downarrow}\rangle |{0}\rangle|\spadesuit\rangle|\Omega_3\rangle 
  \label{psi1b} 
\ee
and
\be
   |\Psi_2\rangle
   = |O\rangle |{\uparrow}\rangle |{-}\rangle|\spadesuit\rangle|\Omega_4\rangle 
  +  |O\rangle |{\downarrow}\rangle |{0}\rangle|\heartsuit\rangle|\Omega_5\rangle
  +  |O\rangle |{\downarrow}\rangle |{+}\rangle|\diamondsuit\rangle|\Omega_6\rangle .
  \label{psi2b}
\ee
From the form of $|\Psi_2\rangle$, in particular the first term in the superposition, it is easy to see that
\be
  P({\uparrow}|\Psi_2) = P(-|\Psi_2) = P(\spadesuit|\Psi_2).\label{firstone}
\ee
From treating different factors as parts of the environment, we also derive
\bea
  P({\uparrow}|\Psi_1) &=& P({\uparrow}|\Psi_2),\\
  P(-|\Psi_1) &=& P(-|\Psi_2),\\
  P(\spadesuit|\Psi_1) &=& P(\spadesuit|\Psi_2).\label{lastone}
\eea
From equations (\ref{firstone})--(\ref{lastone}), we can safely conclude that each of the three branches represented in (\ref{psi1b}) have equal probability, one-third each. Since $|\Psi_1\rangle$ is related to the original $|\Psi\rangle$ by a unitary transformation on the environment, $U_E^{(1)}U_E^{(0)}$, the ESP implies 
\be
  P({\uparrow}|\Psi) = \frac{1}{2}P({\downarrow}|\Psi) = \frac{1}{3}.
\ee
This is precisely the Born Rule prediction for this particular case of unequal amplitudes. The spin-down component of the original state was greater than the spin-up component by a factor of $\sqrt{2}$, and ends up with twice the probability. Other possibilities follow by straightforward extension of the above method. Admittedly, this reasoning only strictly applies when the ratio of different amplitudes is the square root of a rational number; however, since this is a dense set, it seems reasonable to conclude that the Born Rule is established.

This route to the Born Rule has a simple physical interpretation. Take the wave function and write it as a sum over orthonormal basis vectors with equal amplitudes for each term in the sum (so that many terms may contribute to a single branch). Then the Born Rule is simply a matter of counting -- every term in that sum contributes an equal probability.

\section{Mixed Uncertainties and the Multiverse}\label{mixed}

One advantage of the ESP approach to dealing with self-locating uncertainty is that it justifies perfect indifference in the classical case (an agent should give equal credence to being any one of a fixed number of identical observers) as well as the Born Rule in quantum mechanics (see \cite{sc2013}). Furthermore, these results generalize straightforwardly to the mixed case: several quantum branches, any one of which might contain many identical observers. This extravagant-seeming possibility arises quite naturally in the context of the cosmological multiverse, where the decay of a false vacuum during inflation is a quantum event that can create new branches of the wave function with a potentially infinite number of observers.  Page has recently argued that the prospect of classical self-locating uncertainty in large universes poses a crisis for quantum mechanics, as the Born Rule becomes insufficient for calculating the probability of measurement outcomes \cite{Page:2009qe,Page:2009mb,Page:2010bj,Aguirre:2010rw,Albrecht:2012zp}. Our approach provides a unified treatment of classical and quantum self-locating uncertainties, defusing the would-be crisis.

Here we give a simplified version of the discussion in \cite{sc2013}. Consider a quantum state representing a cosmological multiverse $U$, with a set of observers $\{\mathcal{O}_i\}$ in indistinguishable circumstances. The observers may be at different places, different times, and different branches of the wave function. Let the amplitude for the branch containing observer $\mathcal{O}_i$ be denoted $\psi_i$. Then we can (non-trivially) show that ESP implies we should assign such an observer a weight $w_i = |\psi_i|^2$, and the probability that an agent should assign to being observer $\mathcal{O}_i$ is given by
\be
  P(\mathcal{O}_i|U) = \frac{w_i}{\sum_j w_j}.
\ee
The denominator $\sum_j w_j$ will not in general equal unity, since the weights are calculated at different times and possibly for many observers on a single branch. In words, this rule tells us to assign each identical observer in the multiverse a weight given by the amplitude-squared of the branch of the wave function on which they live.

It is then clear how such an observer should describe quantum probabilities. Imagine that each of the identical observers $\mathcal{O}_i$ are going to measure the $z$-component of the spin of a qubit $\sigma_i$, and each qubit is in potentially a different state:
\be
  |\sigma_i\rangle = \alpha_i|{\uparrow}\rangle + \beta_i|{\downarrow}\rangle.
\ee
Such observers find themselves faced with both classical and quantum uncertainty about the probability of the outcome for measuring the spin as either up or down. Here, ESP provides an unambiguous prescription; the probability of observing spin-up is simply given by
\be
  P({\uparrow}|U) = \sum_i |\alpha_i|^2 P(\mathcal{O}_i|U).
\ee
This approach clearly has implications for the cosmological measure problem, allowing us to correctly account for the amplitude of a given semiclassical spacetime geometry as well as for the number of observers within it.

\section{Discussion}

We have proposed that self-locating uncertainty is generic in the process of quantum measurement, and that a proper treatment of such uncertainty leads us directly to the Born Rule \cite{sc2013}. In spirit our approach is similar to that of Vaidman \cite{vaidman2011}, although we have carried the program through in more explicit
detail. The result has the virtue of being relatively physically transparent. The wave function of the universe branches, and initially you don't know which branch you are on; close investigation reveals that the only rational way to apportion credence to the different possibilities is to use the Born Rule.

Formally, our derivation bears a close resemblance to the envariance program of Zurek \cite{zurek2005}, although we believe there are some conceptual advantages. Most importantly, while envariance helps us understand why the Born Rule is a sensible prescription if one thinks of EQM as a probabilistic theory at all, our emphasis on
self-locating uncertainty provides a direct explanation for how such probabilities can arise in a perfectly deterministic theory. In a fundamentally stochastic theory, one thinks of probability as the answer to a question of the form ``how likely is it that this particular outcome will occur?" That philosophy fails in EQM, where it is clear 
that \emph{every} outcome with nonvanishing support in the wave function will occur (in some branch) with probability one. The Born Rule does not tell you the probability that you will end up as ``the observer who measures spin up" (for example); rather, you know with certainty that you will evolve into multiple observers with different
eventual experiences. In our approach, the question is not about which observer you will end up as; it is how the various future selves into which you will evolve should apportion their credences. Since every one of them should use the Born Rule, it is justified to talk \emph{as if} future measurement outcomes simply occur with the
corresponding probability. It is the journey from perfect knowledge to inevitable self-locating uncertainty that is the basis of probability talk in quantum mechanics.

Another advantage of our approach is that it provides a unified framework in which to discuss classical and quantum self-locating uncertainty. This has become an important issue in modern cosmology, in which models of the universe very often predict ``large'' spacetimes with multiple copies of various observers. Our formalism only provides unambiguous guidance in cases where the number of classical observers is finite, so it does not directly address the cosmological measure problem as it appears in models of eternal inflation -- but it seems reasonable that getting the finite case right is an important step towards understanding the infinite case.

\section{Acknowledgements} 

Sean Carroll feels that it has been an honor and a pleasure to take part in the celebration of Yakir Aharonov's 80th birthday and would like to thank Jeff Tollaksen and the organizers of a very stimulating meeting. His work was supported in part by the U.S. Department of Energy, the National Science Foundation, and the Gordon and Betty Moore Foundation.  Charles Sebens's work was supported by the National Science Foundation Graduate Research Fellowship under Grant No. DGE 0718128.

\bibliographystyle{utphys}

\bibliography{bornrule2}

\providecommand{\href}[2]{#2}\begingroup\raggedright\begin{thebibliography}{10}

\bibitem{sc2013}
C.~T. Sebens and S.~M. Carroll, ``{Self-Locating Uncertainty and the Origin of
  Probability in Everettian Quantum Mechanics},''
\href{http://arxiv.org/abs/1405.7577}{{\ttfamily arXiv:1405.7577 [quant-ph]}}.
%%CITATION = ARXIV:1405.7577;%%.

\bibitem{everett}
H.~Everett, ``{``Relative State'' Formulation of Quantum Mechanics},''
  \href{http://dx.doi.org/10.1103/RevModPhys.29.454}{{\em Rev. Mod. Phys.}
  {\bfseries 29} (1957) 454--462}.
  \url{http://link.aps.org/doi/10.1103/RevModPhys.29.454}.

\bibitem{wallace2012}
D.~Wallace, {\em {The Emergent Multiverse: Quantum Theory According to the
  Everett Interpretation}}.
\newblock Oxford University Press, 2012.

\bibitem{vaidman2011}
L.~Vaidman, ``{Probability in the Many-Worlds Interpretation of Quantum
  Mechanics},'' in {\em {The Probable and the Improbable: Understanding
  Probability in Physics, Essays in Memory of Itamar Pitowsky}}, Y.~Ben-Menahem
  and M.~Hemmo, eds.
\newblock Springer, 2011.
\newblock \url{http://philsci-archive.pitt.edu/8558/}.

\bibitem{zurek2005}
W.~Zurek, ``{Probabilities from Entanglement, Born’s Rule $p_ k=\left| \psi_
  {k}\right|^2$ From Envariance},'' {\em Physical Review A} {\bfseries 71}
  no.~5, (2005) 052105.

\bibitem{hartle1968}
J.~Hartle, ``{Quantum Mechanics of Individual Systems},''
  \href{http://dx.doi.org/10.1119/1.1975096}{{\em American Journal of Physics}
  {\bfseries 36} (1968) 704--712}.

\bibitem{Farhi:1989pm}
E.~Farhi, J.~Goldstone, and S.~Gutmann, ``{How Probability Arises in Quantum
  Mechanics},''
\href{http://dx.doi.org/10.1016/0003-4916(89)90141-3}{{\em Annals Phys.}
  {\bfseries 192} (1989) 368}.
%%CITATION = APNYA,192,368;%%.

\bibitem{Aguirre:2010rw}
A.~Aguirre and M.~Tegmark, ``{Born in an Infinite Universe: A Cosmological
  Interpretation of Quantum Mechanics},''
  \href{http://dx.doi.org/10.1103/PhysRevD.84.105002}{{\em Phys.Rev.}
  {\bfseries D84} (2011) 105002},
\href{http://arxiv.org/abs/1008.1066}{{\ttfamily arXiv:1008.1066 [quant-ph]}}.
%%CITATION = ARXIV:1008.1066;%%.

\bibitem{deutsch1999}
D.~Deutsch, ``{Quantum Theory of Probability and Decisions},'' {\em Proceedings
  of the Royal Society of London} {\bfseries A458} (1999) 3129--37.

\bibitem{wallace2002}
D.~Wallace, ``{Quantum Probability and Decision Theory, Revisited},''
  \href{http://arxiv.org/abs/quant-ph/0211104}{{\ttfamily
  arXiv:quant-ph/0211104 [quant-ph]}}.

\bibitem{greaves2004}
H.~Greaves, ``{Understanding Deutsch's Probability in a Deterministic
  Multiverse},'' {\em Studies In History and Philosophy of Science Part B:
  Studies In History and Philosophy of Modern Physics} {\bfseries 35} no.~3,
  (2004) 423--456.

\bibitem{wallace2010b}
D.~Wallace, ``{How to Prove the Born Rule},'' in {\em {Many Worlds?: Everett,
  Quantum Theory, \& Reality}}, S.~Saunders, J.~Barrett, A.~Kent, and
  D.~Wallace, eds., pp.~227--263.
\newblock Oxford University Press, 2010.

\bibitem{gleason}
A.~Gleason, ``{Measures on the Closed Subspaces of a Hilbert Space},'' {\em
  Journal of Mathematics and Mechanics} {\bfseries 6} (1957) 885--894.

\bibitem{lewis1979}
D.~Lewis, ``{Attitudes \textit{de dicto} and \textit{de se}},'' {\em The
  Philosophical Review} {\bfseries 88} no.~4, (1979) 513--543.

\bibitem{Banks:2000fe}
T.~Banks, ``{Cosmological Breaking of Supersymmetry? or Little Lambda Goes Back
  to the Future 2},''
\href{http://arxiv.org/abs/hep-th/0007146}{{\ttfamily arXiv:hep-th/0007146
  [hep-th]}}.
%%CITATION = HEP-TH/0007146;%%.

\bibitem{Carroll:2008yd}
S.~M. Carroll, ``{What if Time Really Exists?},''
\href{http://arxiv.org/abs/0811.3772}{{\ttfamily arXiv:0811.3772 [gr-qc]}}.
%%CITATION = ARXIV:0811.3772;%%.

\bibitem{Boddy:2014eba}
K.~K. Boddy, S.~M. Carroll, and J.~Pollack, ``{De Sitter Space Without Quantum
  Fluctuations},''
\href{http://arxiv.org/abs/1405.0298}{{\ttfamily arXiv:1405.0298 [hep-th]}}.
%%CITATION = ARXIV:1405.0298;%%.

\bibitem{zurek81}
W.~H. Zurek, ``{Pointer Basis of Quantum Apparatus: Into What Mixture Does the
  Wave Packet Collapse?},''
  \href{http://dx.doi.org/10.1103/PhysRevD.24.1516}{{\em Phys. Rev. D}
  {\bfseries 24} (Sep, 1981) 1516--1525}.
  \url{http://link.aps.org/doi/10.1103/PhysRevD.24.1516}.

\bibitem{jooszeh85}
E.~Joos and H.~Zeh, ``{The Emergence of Classical Properties Through
  Interaction with the Environment},''
  \href{http://dx.doi.org/10.1007/BF01725541}{{\em Zeitschrift für Physik B
  Condensed Matter} {\bfseries 59} (1985) 223--243}.
  \url{http://dx.doi.org/10.1007/BF01725541}.

\bibitem{wallace2010c}
D.~Wallace, ``{Decoherence and Ontology},'' in {\em {Many Worlds?: Everett,
  Quantum Theory, \& Reality}}, S.~Saunders, J.~Barrett, A.~Kent, and
  D.~Wallace, eds., pp.~53--72.
\newblock Oxford University Press, 2010.

\bibitem{hartle2007}
J.~Hartle and M.~Srednicki, ``{Are We Typical?},'' {\em Phys. Rev. D}
  {\bfseries 75} no.~12, (2007) 123523.

\bibitem{page2007}
D.~Page, ``{Typicality Defended},''
  \href{http://arxiv.org/abs/arxiv:0707.4169}{{\ttfamily arxiv:0707.4169}}.

\bibitem{srednicki2010}
M.~Srednicki and J.~Hartle, ``{Science in a Very Large Universe},'' {\em
  Physical Review D} {\bfseries 81} no.~12, (2010) 123524.

\bibitem{bostrom2002}
N.~Bostrom, {\em {Anthropic Bias: Observation Selection Effects in Science and
  Philosophy}}.
\newblock Routledge, 2002.

\bibitem{meacham2008}
C.~Meacham, ``{Sleeping Beauty and the Dynamics of \textit{de se} Beliefs},''
  {\em Philosophical Studies} {\bfseries 138} no.~2, (2008) 245--269.

\bibitem{manleyF}
D.~Manley, ``{On Being a Random Sample},'' {\em unpublished} .

\bibitem{elga2004}
A.~Elga, ``{Defeating Dr. Evil with Self-Locating Belief},'' {\em Philosophy
  and Phenomenological Research} {\bfseries 69} no.~2, (2004) 383--396.

\bibitem{Page:2009qe}
D.~N. Page, ``{The Born Rule Dies},''
  \href{http://dx.doi.org/10.1088/1475-7516/2009/07/008}{{\em JCAP} {\bfseries
  0907} (2009) 008},
\href{http://arxiv.org/abs/0903.4888}{{\ttfamily arXiv:0903.4888 [hep-th]}}.
%%CITATION = ARXIV:0903.4888;%%.

\bibitem{Page:2009mb}
D.~N. Page, ``{Born Again},''
\href{http://arxiv.org/abs/0907.4152}{{\ttfamily arXiv:0907.4152 [hep-th]}}.
%%CITATION = ARXIV:0907.4152;%%.

\bibitem{Page:2010bj}
D.~N. Page, ``{Born's Rule Is Insufficient in a Large Universe},''
\href{http://arxiv.org/abs/1003.2419}{{\ttfamily arXiv:1003.2419 [hep-th]}}.
%%CITATION = ARXIV:1003.2419;%%.

\bibitem{Albrecht:2012zp}
A.~Albrecht and D.~Phillips, ``{Origin of Probabilities and Their Application
  to the Multiverse},''
\href{http://arxiv.org/abs/1212.0953}{{\ttfamily arXiv:1212.0953 [gr-qc]}}.
%%CITATION = ARXIV:1212.0953;%%.

\end{thebibliography}\endgroup
\end{document}